\documentclass[aps,amssymb,amsmath, twocolumn, prb, superscriptaddress]{revtex4}
\usepackage{amsmath}    
\usepackage{bbm}    
\usepackage{amsfonts}
\usepackage[colorlinks, citecolor=red]{hyperref}
\usepackage{hyperref}
\usepackage{float}
\usepackage{color}
\usepackage{epsfig}
\usepackage[up]{subfigure}
\usepackage{epstopdf}
\newcommand{\be}{\begin{equation}}
\newcommand{\ee}{\end{equation}}
\newcommand{\bc}{\begin{center}}
\newcommand{\ec}{\end{center}}
\newcommand{\bea}{\begin{eqnarray}}
\newcommand{\eea}{\end{eqnarray}}
\newcommand{\ba}{\begin{array}}
\newcommand{\ea}{\end{array}}

\newcommand{\norm}[1]{\lvert #1 \rvert}

\usepackage{ulem}

\begin{document}

\title{Entanglement Properties of Localized States in 1D Topological Quantum Walks}

\author{C.M.~Chandrashekar}
\email{c.madaiah@oist.jp}
\affiliation{Quantum Systems Unit, Okinawa Institute of Science and Technology Graduate University, Okinawa, Japan}
\author{H.~Obuse}
\email{hideaki.obuse@eng.hokudai.ac.jp}
\affiliation{Department of Applied Physics, Hokkaido University, Sapporo 060-8628, Japan}
\author{Th.~Busch}
\email{thomas.busch@oist.jp}
\affiliation{Quantum Systems Unit, Okinawa Institute of Science and Technology Graduate University, Okinawa, Japan}

%================================================

\begin{abstract}
\noindent
The symmetries associated with discrete-time quantum walks (DTQWs) and the flexibilities in controlling their dynamical parameters allow to create a large number of topological phases. An interface in position space, which separates two regions with different topological numbers, can, for example, be effectively modelled using different coin parameters for the walk on either side of the interface. Depending on the neighbouring numbers, this can lead to localized states in one-dimensional configurations and here we carry out a detailed study into the strength of such localized states. We show that it can be related to the amount of entanglement created by the walks, with minima appearing for strong localizations. This feature also persists in the presence of small amounts of $\sigma_x$ (bit flip) noise.  
\end{abstract}

\maketitle
%==========================
\section{Introduction}
\label{intro}
%==========================

Quantum walks\,\cite{Ria58+, DM96} can be used to efficiently create non-classical states and  have therefore been of large interest for designing quantum algorithms\,\cite{Amb03, CCD+03, Sze04, AA05, MN07, Amb07, MNR+12} and  realizing universal quantum computation\,\cite{Chi09, LCE10}. However, in recent years, quantum walks have also been employed to understand the dynamics of a considerable range of other physical processes, for example dielectric breakdown in driven electron system\,\cite{OKA05}, transport in biological or chemical systems\,\cite{ECR07, MRL08, PH08} or effects in relativistic quantum dynamics\,\cite{ DM96, Str07, CBS10, GDB12, Cha13}. One of the more recent topics of interest is the creation of topological phases using quantum walks. 

Topological properties of materials have recently been recognised as a rich source of interesting physics and have led to a new class  known as topological insulators (TIs)\,\cite{HK10, QZ11, Moo10}. However, only a small number of natural TIs are known and therefore interest in creating artificial materials with non-trivial topological states is a prime research activity. Discrete-time quantum walks (DTQWs) are a method to create such states, as they can simulate time-independent lattice Hamiltonians with the required symmetries. At the same time they possess additional degrees of freedom, for example the possibility for varying coin operations, which can lead to much richer system\,\cite{KRB10, OK11, KBF12, Asb12, AO13, TAD14}. Progress in the theoretical understanding of these systems is going hand in hand with current advances in experimental implementations and engineering of quantum walks in various physical systems\cite{DLX03+}.  Exploring topological phases using DTQWs has therefore emerged as a promising approach to realizing TIs in artificial materials. 
 
The nontrivial topological phases of TIs are intricately linked to the presence or absence of certain symmetries, namely, time-reversal symmetry, particle-hole symmetry, and chiral symmetry\cite{Schnyder08}. For one-dimensional DTQWs with the all three symmetries present (belonging to class BDI), the topological properties have recently been studied using a split-step\,\cite{KRB10} and a double split-step DTQW\,\cite{AO13}. Due to the $2\pi$ periodicity of the quasi-energies, the topological numbers of a 1D DTQW are defined not only for $0$ but also for $\pi$ quasi-energies, which means that they become $\mathbbm{Z \times Z}$ winding numbers\,\cite{AO13}. 
Consequently, at the interface where two domains with different winding numbers are connected, topologically protected surface states appear at the two specific quasi-energies. Because of the one-dimensionality and the particle-hole symmetry, these surface states are the localized Majorana edge states,  which have recently been experimentally observed\,\cite{KBF12}.

As the winding number is a function of the angle $\theta_i$ used in the quantum coin operation for each split-step, the phase diagram of a TI can be written in terms of the angle on either side of the interface. This allows to identify the combinations that lead to the appearance of localized states at the interface, but does not give any information about the strength of the localization, i.e.~the probability of finding the particle at the interface. While for some configurations the localization is very strong and only a small probability of finding the particle away from the interface exist, for other configurations it can be weaker.  Knowing which configurations result in strongly topologically localized states is necessary to identify parameters that lead to TIs with a strongly insulating bandstructure. 

In the following we will show that strong localization at the interface due to topological effects can be signaled by a minimum in entanglement generated during  split-step and double split-step DTQWs. This is in contrast to the properties found for localized states due to disordered coin operations in DTQWs, where for the standard DTQW an enhancement of entanglement is seen for temporal\,\cite{Cha12, VAR13} and spatio-temporal disorder and only a small decrease is seen for purely spatial disorder\,\cite{Cha12}. In addition, we will also discuss the effect of noise on topologically localized states and show that they are robust against $\sigma_x$ (bit flip) noise.

 %=========================================
\section{Topological Quantum Walks}
\label{TQW}
 %=========================================
A 1D DTQW is defined for a system composed of a particle space and a position space. The basis states of the particle space can be any two internal states represented by $|0\rangle$ and $|1\rangle$ and the basis states of the position space are defined on $|x\rangle$, where $x$ is an integer. If the initial state is given by a particle in state
 $|\Psi_\text{in}\rangle = (\alpha|0\rangle + \beta |1\rangle )\otimes |x = 0\rangle$, which is located at the origin, each step of the walk is composed of a quantum coin operation 
  \begin{equation}
  \label{eq:1}
  R_{\theta}  \equiv   \begin{bmatrix}
  \cos \left(\theta/2 \right)      &       - \sin \left( \theta/2 \right )\\
  \sin \left (\theta/2 \right )  &  \quad\cos \left (\theta/2 \right )
                                  \end{bmatrix} \otimes \mathbb{I},
  \end{equation}
followed by a position shift operation
\be
\label{eq:shift}
S = |0\rangle \langle 0 | \otimes |x-1\rangle \langle x |    + |1 \rangle \langle 1 | \otimes |x+1 \rangle \langle x |.
\ee
The unitary operator $W(\theta) = S R_{\theta}$ therefore defines one step of the standard DTQW and the state after $t$ steps is given by $|\Psi_t \rangle =
[SR_{\theta}]^t |\Psi_\text{in}\rangle$.

The eigenstates of the single time step operator can be written as
\begin{equation}
   W(\theta) |\psi_\varepsilon \rangle = e^{-i\varepsilon} |\psi_\varepsilon \rangle,
\end{equation}
where the quasi-energies $\varepsilon$ are real and have $2 \pi$ periodicity. These spectral properties of $W(\theta)$ give insight into the long time behavior of the walk and therefore also the behavior of topologically protected localized states.

Nontrivial topological phases in DTQW can be found when the evolution operator $W(\theta)$ indicates the presence of certain specific symmetries, such as time-reversal, particle-hole, or chiral symmetries. For 1D  systems particle-hole or chiral symmetries are known to be important\,\cite{Schnyder08} to lead to two different topological numbers, $\nu_0$ and $\nu_\pi$, for the quasi-energies $\varepsilon=0$ and $\pi$, which in turn leads to edge states that are localized at the interfaces across which the topological numbers change\,\cite{AO13}.

As all the matrix elements of the operator $W(\theta)$ used for defining the DTQW above are real, particle-hole symmetry is automatically guaranteed.
To ensure chiral symmetry for $W(\theta)$ one needs to ensure the existence of a chiral operator $\Gamma$ which satisfy the relation  
\begin{equation}
 \Gamma W(\theta) \Gamma^{-1} = W(\theta)^{-1},
\label{eq:chiral}
\end{equation}
For this we will first decompose the operator $W(\theta)$ as
\begin{equation}
W(\theta) = F\cdot G,
\end{equation}
where $F$ and $G$ are two sub-steps with each being a composition of coin ($R_{\theta}$) and shift operator ($S$). They are related by
\begin{equation}
\Gamma F \Gamma^{-1} = G^{-1},
\label{eq:FG}
\end{equation}
and the above expression is guaranteed if the components of both, $F$ and $G$, satisfy
\begin{equation}
\Gamma R_\theta \Gamma^{-1} = R_\theta^{-1}\qquad\text{and}\qquad
\Gamma S \Gamma^{-1} = S^{-1}.
\label{eq:chiralSR}
\end{equation}
This leads to a chiral symmetry operator of the form 
\begin{equation}
\Gamma \equiv  
\sigma_x \otimes \mathbb{I},\quad\text{with}\quad
\sigma_x=
 \begin{bmatrix} \begin{array}{cc}
  0     &   1  \\
  1 & 0 \end{array} \end{bmatrix}.
\label{eq:chiral symmetry operator}
\end{equation}

The topological numbers ($\nu_0, \nu_{\pi}$) of the 1D DTQW stemming from this kind of chiral symmetry have been calculated recently\,\cite{AO13}. If $W(\theta)$ satisfies chiral symmetry (Eq.\,(\ref{eq:chiral})), a counterpart state with opposite sign of the quasi-energy is guaranteed, 
\begin{equation}
 W(\theta) |\psi_{-\varepsilon}\rangle = e^{+i\varepsilon} |\psi_{-\varepsilon}\rangle,
\end{equation}
where 
\be
|\psi_{-\varepsilon}\rangle \equiv \Gamma
|\psi_\varepsilon\rangle.
\ee
Taking into account the $2\pi$ periodicity of $\varepsilon$, the above relation for the edge states at $\varepsilon = 0$ and  $\pi$ is therefore identical to the eigenstate equation of the chiral symmetry operator $\Gamma$,
\begin{equation}
 \Gamma |\psi_{0,\pi}\rangle = \pm |\psi_{0,\pi}\rangle,
\label{eq:chiral edge states}
\end{equation}
with the eigenvalues $\pm 1$. 
\par
In the following we will focus on  two specific DTQW with chiral symmetry and discuss their topological properties. First we consider a DTQW with each step split into two with different coin parameters $\theta_i$\,\cite{KRB10} as
\be
 \label{eq:4}
 W(\theta_1, \theta_2) =    S_{+}R_{\theta_2}S_{-}R_{\theta_1},
\ee
and for which the position split shift operators are 
\bea
S_{-} = |0\rangle \langle 0 | \otimes |x-1\rangle \langle x |    + |1 \rangle \langle 1 | \otimes |x \rangle \langle x |  ;
\eea
\bea
S_{+} = |0 \rangle \langle 0 | \otimes |x \rangle \langle x | +  |1\rangle \langle 1 | \otimes |x+1\rangle \langle x |. 
\eea
\par
\begin{figure}[tb]
   \includegraphics[width=.98\linewidth]{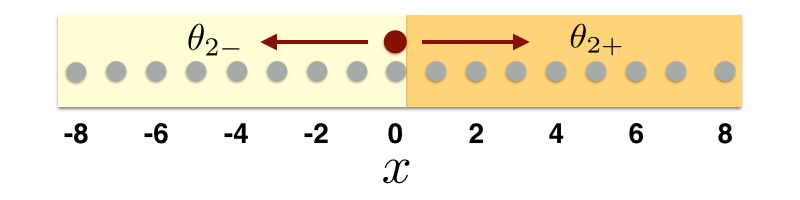}
  \caption{Schematic of the position space showing the boundary created by $\theta_{2-}$ and $\theta_{2+}$, which leads to the appearance of topological properties.}
  \label{schematicTQW}
\end{figure}

 
To create a real space boundary between topologically distinct phases and reveal non-trivial topological properties at the interface, one can choose different $\theta_{2}$ to the left ($R_{\theta_{2-}}$) and right side ($R_{\theta_{2+}}$) of a point in the position space as shown in Fig.~\ref{schematicTQW}, while
defining the coin operation $R_{\theta_1}$ uniformly on the entire position space.

\begin{figure}[tb]
    \includegraphics[width=\linewidth]{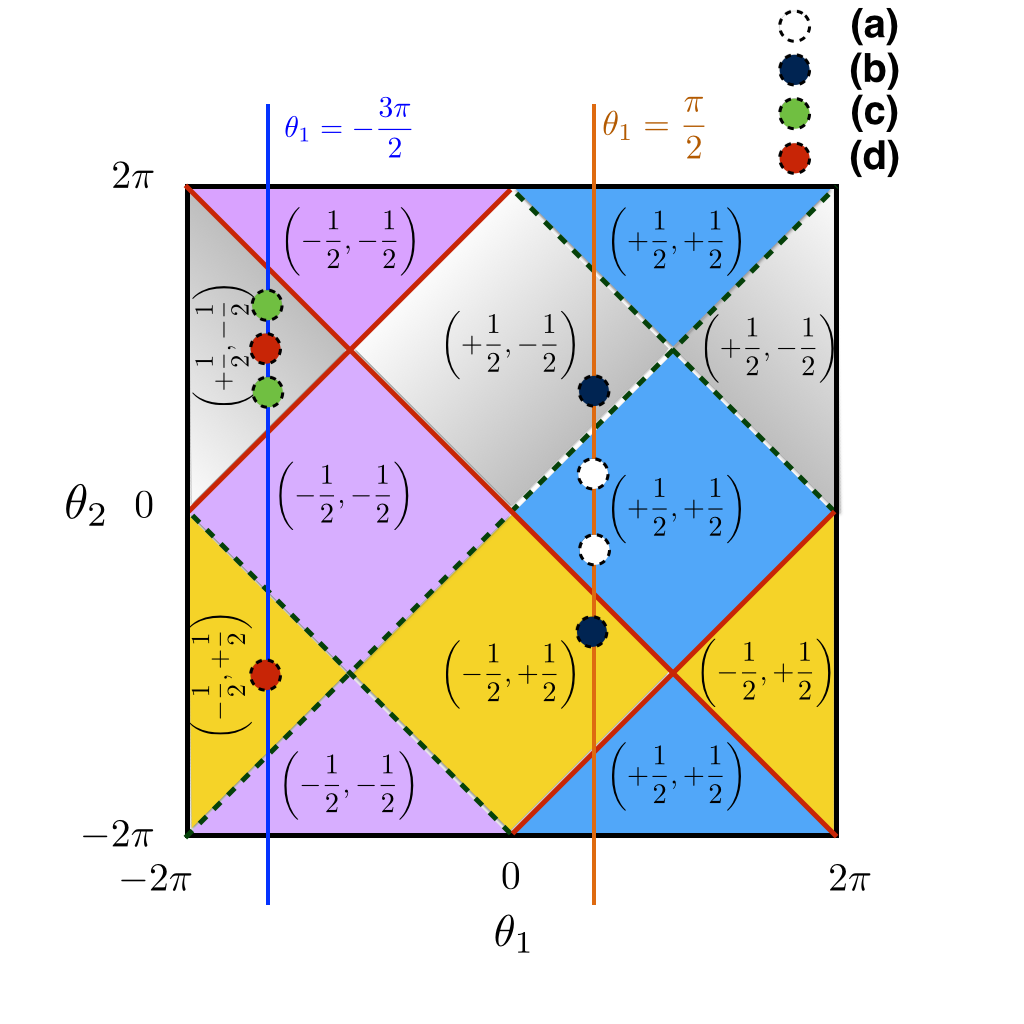}
  \caption{Phase diagram of the topological numbers ($\nu_0, \nu_\pi$) associated with the split-step DTQW as a function of the coin parameters $\theta_1$ and $\theta_2$.  Topologically distinct gapped phases are separated by lines where the gap closes at either $\varepsilon = 0$ (solid lines) or $\varepsilon =\pi$ (dashed lines).  Fours pairs of circles with different colours and labeled as (a), (b), (c) and (d) are marked to identify different parameters corresponding to the probability distributions shown in Fig.~\ref{fig2:ProbDist}.}
\label{fig:Phasediagram}
 \end{figure}

The topological numbers $(\nu_0, \nu_\pi)$ for this split-step DTQW as a function of the coin parameters $\theta_1$ and $\theta_2$ are shown in Fig.~\ref{fig:Phasediagram}\,\cite{Asb12}. Using this phase diagram one can identify combinations of $\theta_{2-}$ and $\theta_{2+}$ that are located in regions with different topological numbers and therefore lead to an interface where a localized state can exist.

\begin{figure}[tb]
   \includegraphics[width=\linewidth]{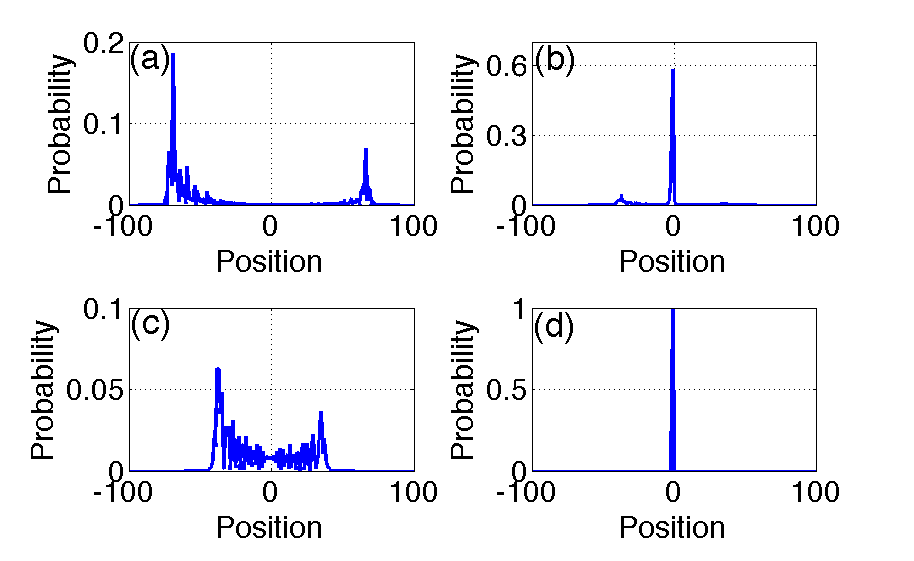}
  \caption{Probability distributions   of the  split-step DTQW after 100 steps with a particle initially in the state $|\Psi_\text{in}\rangle = \frac{1}{\sqrt 2} (|0\rangle + |1\rangle ) \otimes |x =0\rangle$.  For (a)  and (b) $\theta_1 = \pi/2$ and   ($\theta_{2-}$,  $\theta_{2+}=(-\pi/4, \pi/4)$ and $(-3\pi/4, 3\pi/4)$, respectively. For (c) and (d)  $\theta_1 =-3\pi/2$   the ($\theta_{2-}$,  $\theta_{2+})=(5\pi/4, 3\pi/4)$ and $(-\pi, \pi)$, respectively. These parameters correspond to the circles in Fig.~\ref{fig:Phasediagram} and are chosen such that in (a) and (c) the $\theta_{2-}$ and  $\theta_{2+}$ are from regions with the same topological invariant, whereas in (b) and (d) they stem from regions with distinct topological numbers.}
 \label{fig2:ProbDist}
  \end{figure}

Examples of the behaviour can be seen in Figs.~\ref{fig2:ProbDist}(b) and \ref{fig2:ProbDist}(d), where we show the spatial probability distribution after 100 steps for $(\theta_{1} ,\theta_{2+},\theta_{2-})=(\pi/2,-3\pi/4, 3\pi/4)$ and  $(-3\pi/2,-\pi, \pi)$. In both cases localization is clearly visible. If, on the other hand, $\theta_{2+}$ and $\theta_{2-}$ correspond to regions with the same topological number, the probability at $x=0$ decreases with time, indicating the absence of a localized state. Examples of this are shown in Figs.~\ref{fig2:ProbDist}(a) and \ref{fig2:ProbDist}(c) for values of $(\theta_{1} ,\theta_{2+},\theta_{2-})=(\pi/2,-\pi/4, \pi/4)$ and $(-3\pi/2,5\pi/4, 3\pi/4)$, respectively.

\begin{figure}[tb]
 \begin{center}
   \includegraphics[width=\linewidth]{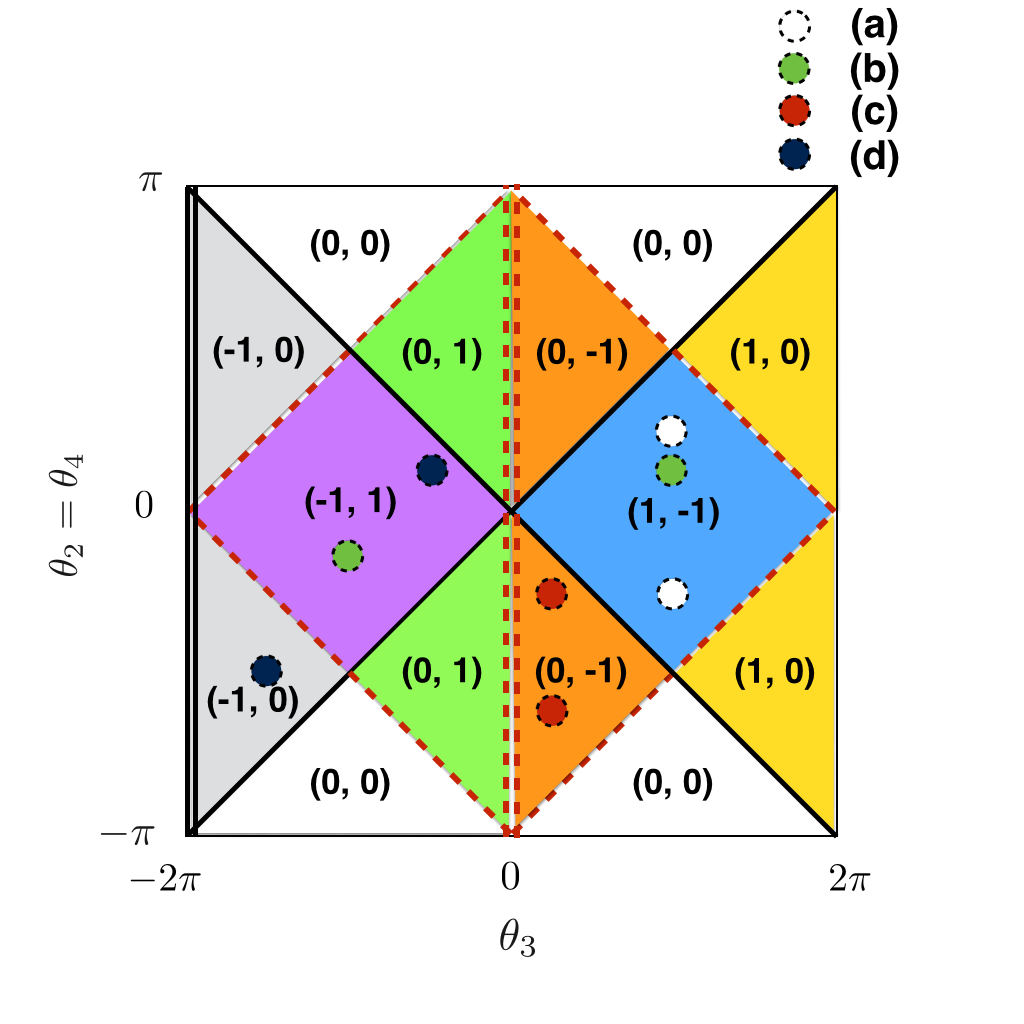}
   \caption{Phase diagram of the topological numbers associated with double split-step DTQW as a function of the coin parameters $\theta_2$ and $\theta_3$. The parameters $\theta_2$ and $\theta_4$ are set equal to ensures CS and we have chosen $\theta_1= 0$. Topologically distinct gapped phases are separated by lines where the gap closes at either $\varepsilon = 0$ (solid lines) or $\varepsilon = \pi$ (dashed lines).  Four pairs of circles with different shades labeled as (a), (b), (c) and (d) are marked to identify the different parameters used to generate probability distributions shown in Fig.~\ref{fig4:ProbDist4step}.}
\label{fig3:phasediagram}
 \end{center}
 \end{figure}

\begin{figure}[tb]
   \includegraphics[width=\linewidth]{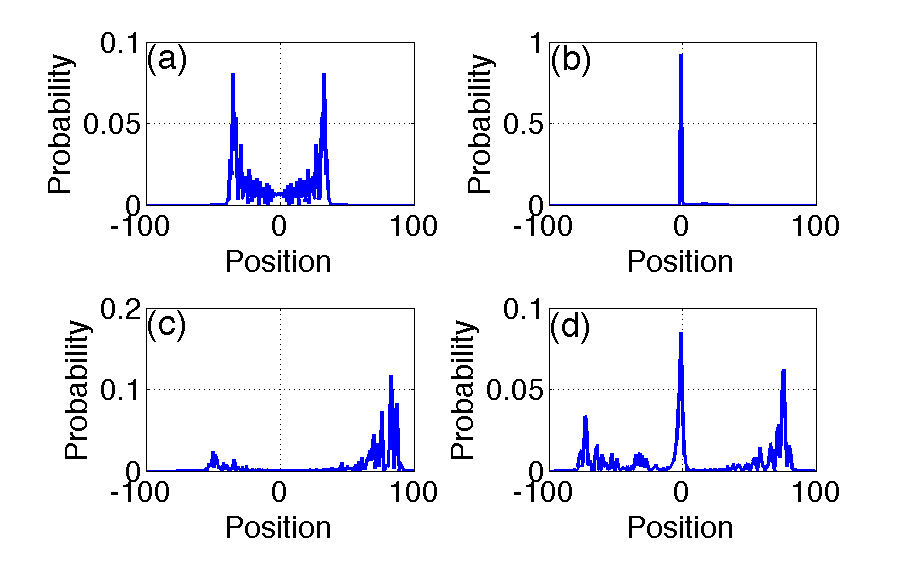}
  \caption{Probability distributions of the double split-step DTQW after 50 steps of the walk for a particle with the initial state $|\Psi_\text{in}\rangle =
  \frac{1}{\sqrt 2} (|0\rangle + |1\rangle ) \otimes |x =0\rangle$. The chosen parameters for   $(\theta_{2-}, \theta_{2+} ; \theta_{3-},
  \theta_{3+})$ are marked with circles in Fig.~\ref{fig3:phasediagram}, (a) $(-\pi/4, \pi/4 ; \pi, \pi)$ (b) $(-\pi/8, \pi/8 ; -\pi, \pi)$, (c) $(-\pi/4, -3\pi/4 ; \pi/4, \pi/4)$ and (d) $(-3\pi/8, \pi/8 ; -3\pi/2, -\pi/2)$.
  The parameters in (a) and (c) are from regions with the same topological number and the absence of a localized state is evident, whereas the parameters in (b) and (d) are from regions with different topological numbers, which leads to localization.}
 \label{fig4:ProbDist4step}
  \end{figure}

A second class of  DTQWs with rich topological features are double split-step evolutions\,\cite{ AO13}. These are described by four parameters $\theta_i$, leading to an effective Hamiltonian with long range hopping that results in higher values for  winding numbers and  topological numbers.
Each step in the double split-step walk is a composition of the operators
\be
 W(\theta_1, \theta_2, \theta_3, \theta_4) =  S_{+}R_{\theta_4}S_{+}R_{\theta_3}S_{-}R_{\theta_2}S_{-}R_{\theta_1},
 \ee
where setting $\theta_2 = \theta_4$ ensures chiral symmetry (CS). For simplicity we will also set $\theta_1 = 0$ in the following and in Fig.~\ref{fig3:phasediagram}  we show the phase diagram as function of $\theta_2$ and $\theta_3$. Regions with different topological numbers can again be clearly identified \cite{AO13} and in Fig.~\ref{fig4:ProbDist4step} examples of spatial probability distributions for the situation in which different coin parameters have been used on the left and the right of the initial position, creating a boundary at the origin, are shown. The four pairs of parameters $(\theta_2,\theta_3)$ used to generate these probability distributions are marked with circles of different color in the phase diagram (Fig.~\ref{fig3:phasediagram}).  For the parameters$(\theta_{2-}, \theta_{2+};\theta_{3-}, \theta_{3+}) = (-\pi/8, \pi/8 ; -\pi, \pi)$ and $(-3\pi/8, \pi/8 ;-3\pi/2, -\pi/2)$, chosen from regions with
 different topological numbers, the probabilities of finding the particle at $x=0$ remains high,
 indicating the presence of localized state (see Figs.~\ref{fig4:ProbDist4step}(b) and \ref{fig4:ProbDist4step}(d)). 
For the parameters $(\theta_{2-}, \theta_{2+} ; \theta_{3-},\theta_{3+}) = (-\pi/4, \pi/4 ; \pi, \pi)$ and $(-\pi/4, 3\pi/4 ; \pi/4, \pi/4)$, chosen from regions with same topological numbers, the probability of finding the particle at $x=0$ is very low, indicating the dominance of diffusion (see Figs.\,\ref{fig4:ProbDist4step}(a) and \ref{fig4:ProbDist4step}(c)).  One should note that it is possible to generate localized states for certain sets of parameters from the regions with the same topological invariant. Those, however, have energies different from $0$ or $\pi$ and are therefore distinguishable from localized states originating from topological effects.

%==========================
\section{Entanglement properties}
\label{esTQW}
%==========================

DTQW are known to entangle the particle and the position space. The degree of entanglement depends on the parameters that define the evolution operators\,\cite{MK07, GC10} and it is intriguing to explore this quantity for topological quantum walks. While in the previous section localized states were shown to appear when choosing coin parameters for the left and right regions from areas with different topological numbers in the phase diagram, no indication could be drawn from this about the strength of the localized state. As often the probability of the diffusing component can be higher than the one of the localized part, it is important to identify the parameters that lead to the highest probability for finding a strongly topologically localized state at $x=0$, in order to create artificially synthesized TIs.  
In this section we ask and answer the question if entanglement is an effective measure to identify the configurations of parameters that result in strongly localized states.  For this we calculate the entanglement generated by different topological quantum walks and identify the regions which lead to strongly localized states. 

To quantify the entanglement between the particle and the position space we will use negativity, which is the absolute sum of the negative eigenvalues of the partial transpose of the density operator, $\rho = |\Psi_t \rangle \langle \Psi_t |$. It is given by 
\bea
{\mathcal N}(\rho) = \sum_{i}  \frac {\norm{\lambda_i} - \lambda_{i}}{2}
\eea
where the $\lambda_i$ are the eigenvalues of $\rho$.  

\begin{figure}[tb]
 \begin{center}
      \includegraphics[width=\linewidth]{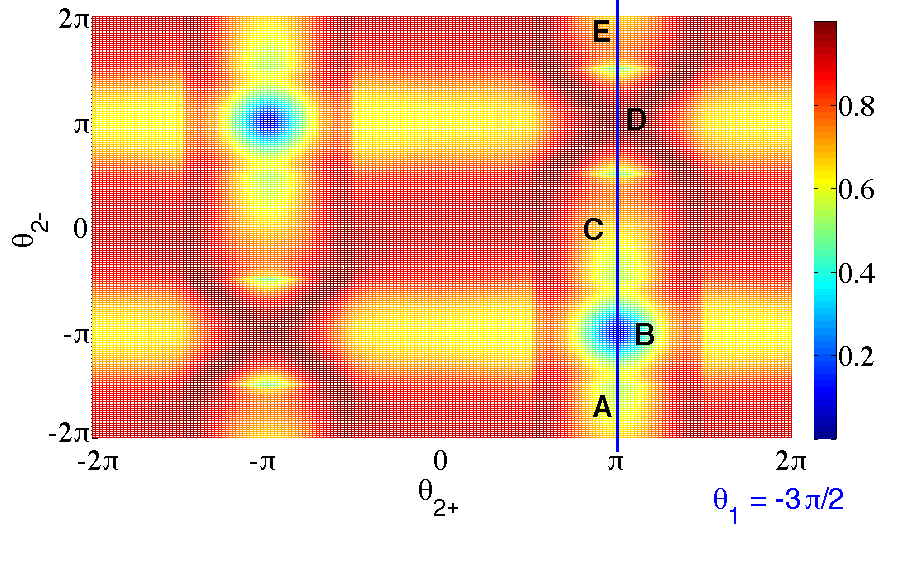}
  \caption{Negativity as a function of $\theta_{2-}$ and $\theta_{2+}$ for the split-step DTQW with $\theta_{1}=-3\pi/2$. The initial state was given by $|\Psi_\text{in}\rangle = \frac{1}{\sqrt 2}(|0\rangle + |1\rangle)\otimes |x=0\rangle$ and the results are shown after 100 steps of walk.
  }
\label{NegProfile2stateA}
 \end{center}
 \end{figure}

To reduce the number of free parameters we initially fix $\theta_1 = -3\pi/2$  for the split-step DTQW and show in Fig.~\ref{NegProfile2stateA}  the negativity as function of  $\theta_{2-}$ and $\theta_{2+}$.  A varied landscape is clearly visible and to interpret the structure, one can map the diagram to the one for the topological numbers.

\begin{figure}[tb]
 \begin{center}
   \includegraphics[width=\linewidth]{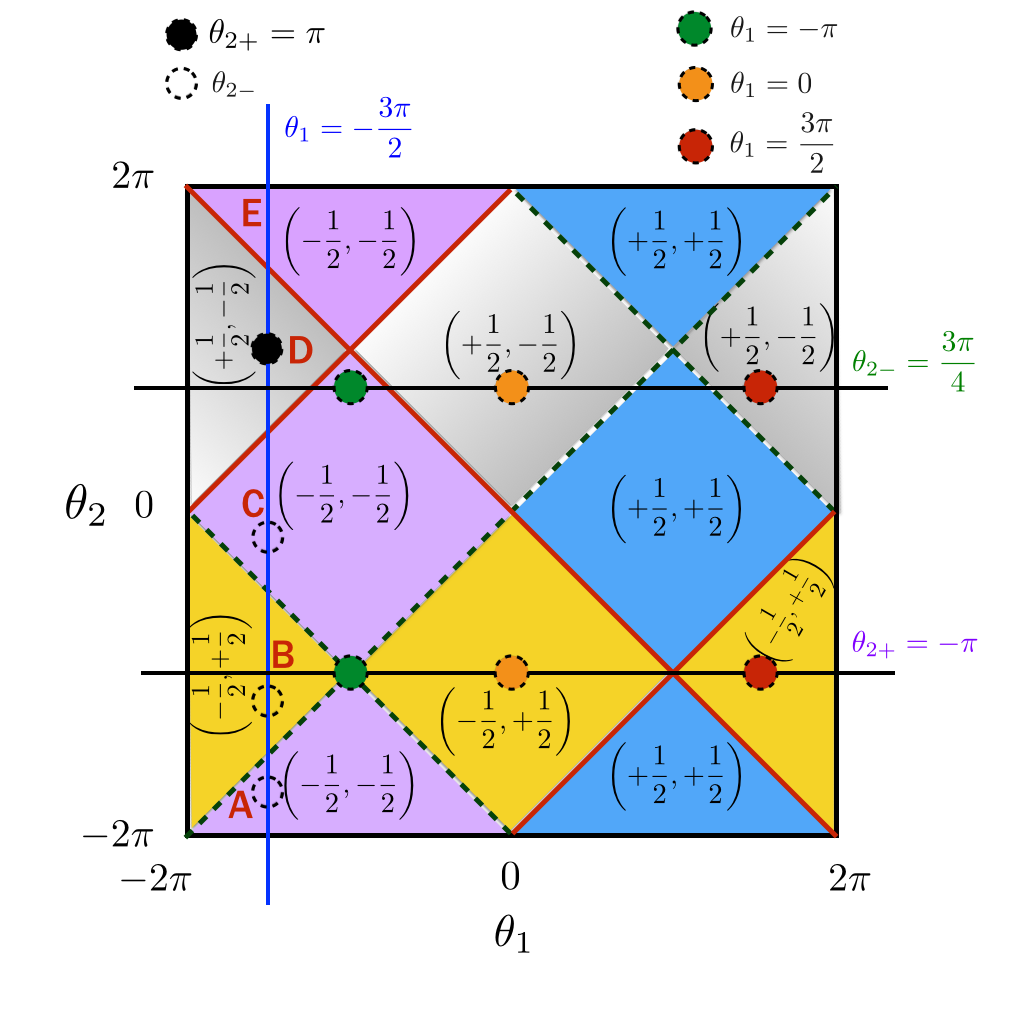}
	  \caption{Phase diagram of the split-step DTQW. The vertical line corresponds to the results shown in Fig.~\ref{NegProfile2stateA} and the horizontal lines corresponds to the negativity presented in Fig.~\ref{spectrumTheta1theta2}.}
  \label{twostatePhaseDia2}
 \end{center}
 \end{figure}

For this we show in Fig.~\ref{twostatePhaseDia2} the phase diagram again and the vertical line at $\theta_1 = -3\pi/2$ indicates the parameters for which the negativity in displayed in Fig.~\ref{NegProfile2stateA}.  If  $\theta_{2+} =\pi$ (marked with a filled circle in region {\bf D}) and $\theta_{2-}$ is ranging from $-2\pi$ to $2\pi$, one can see that the values of negativity have two clear trends. They are high if $\theta_{2-}$ is in a region with the same topological invariant as $\theta_{2+} $(here the gray area around {\bf D})
and low if $\theta_{2-}$ is in a region that has a different topological invariant. In fact, one can see that a minimum appears when $\theta_{2-}$ is from region {\bf B} and we find that in general  lower values of negativity indicate the presence of localized states with smaller fractions of the particle's amplitude diffused. (see Fig.~\ref{fig2:ProbDist}(d)). 

\begin{figure}[tb]
 \begin{center}
   \includegraphics[width=\linewidth]{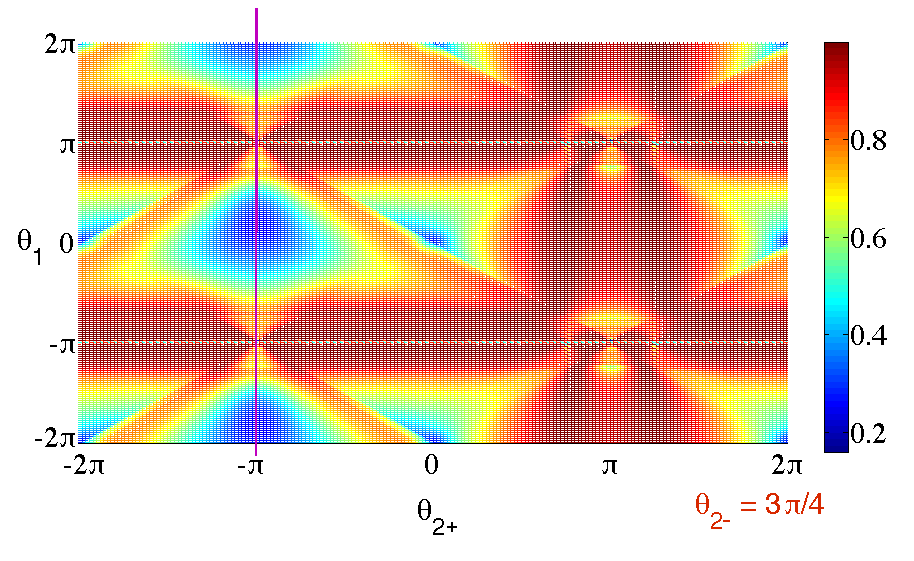}
    \caption{Negativity of the split-step DTQW as a function of $\theta_1$ and $\theta_{2+}$ for  $\theta_{2-} = 3\pi/4$.  The initial state was $|\Psi_\text{in}\rangle = \frac{1}{\sqrt 2}(|0\rangle + |1\rangle)\otimes |x=0\rangle$ and the results shown are after 100 steps of walk.}
 \label{spectrumTheta1theta2}
 \end{center}
 \end{figure}

In Fig.\,\ref{spectrumTheta1theta2}  we show the negativity as a function of $\theta_1$ and $\theta_{2+}$ by fixing $\theta_{2-}=3\pi/4$. When $\theta_{2+} = -\pi$ the values of $\theta_{2-}$ and $\theta_{2+}$ will be in regions with different topological numbers for all values of $\theta_{1}$, except for $\theta_1=\pm \pi$, where regions of different  topological numbers meet.  This can be seen from Fig.\,\ref{twostatePhaseDia2} where the two horizontal lines indicate the topological regions in which $\theta_{2-}$ and $\theta_{2+}$ lie when $\theta_{1}$ is ranging from $-2\pi$ to $2\pi$. The corresponding values of negativity correspond to the vertical line in Fig.~\ref{spectrumTheta1theta2} and one can clearly see low values of negativity for all values of $\theta_1$, except at the points $\theta_{1} = \pm \pi$.  A general comparison of the phase diagram (Fig.~\ref{twostatePhaseDia2}) and the negativity profiles (Figs.~\ref{NegProfile2stateA} and \ref{spectrumTheta1theta2}) for different combinations of $\theta_{1}
 $,  $\theta_{2-}$ and $\theta_{2+}$ shows that low values for the negativity appear whenever the combination of $\theta_{2-}$ and $\theta_{2+}$  is chosen from regions with different topological invariant.  This indicates that a low area in the negativity landscape can be effectively used to identify the combinations that result in localized states, with the minima corresponding to localized states with zero or minimal diffusion component.
\par
\begin{figure}[tb]
 \begin{center}
   \includegraphics[width=\linewidth]{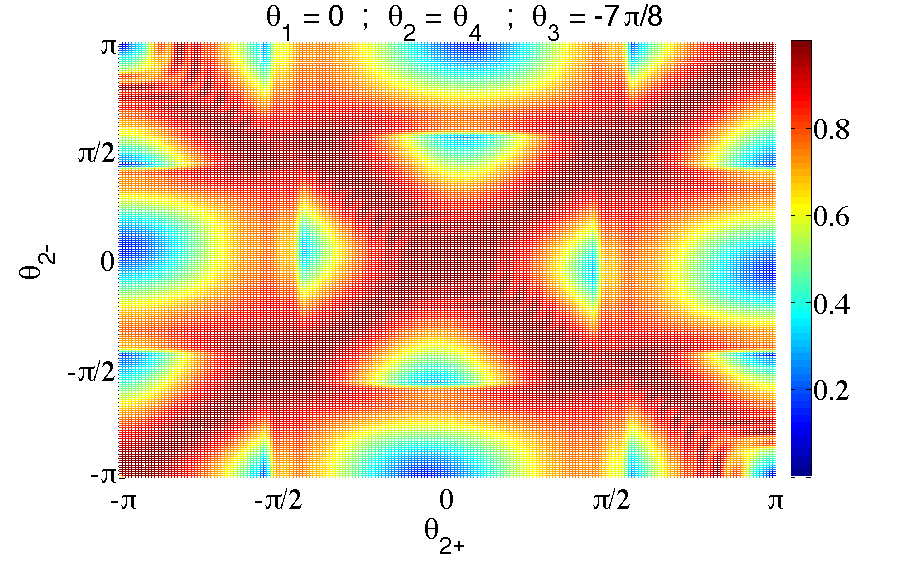}
   \caption{Negativity of the double split-step DTQW as a function of $\theta_{2-}$ and $\theta_{2+}$ when $\theta_1 =0$, $\theta_3 = -7\pi/8$ and $\theta_{4\pm} =\theta_{2\pm}$. 
    The initial state was $|\Psi_\text{in}\rangle = \frac{1}{\sqrt 2}(|0\rangle + |1\rangle)\otimes |x=0\rangle$ and the results shown are after 50 steps of walk.
   }
\label{negprofile4step}
 \end{center}
 \end{figure}

Similarly, a negativity plot for the double split-step DTQW can be effectively used to identify the combination of parameters which lead to strongly localized states. In Fig.~\ref{negprofile4step} we show the negativity as a function of $\theta_{2+}$ and $\theta_{2-}$ for $\theta_{1}=0$, $\theta_{3} = -7\pi/8$ and $\theta_{4\pm} = \theta_{2\pm}$.  The visible valleys of entanglement corresponds to parameter ranges in which a strongly localized state is obtained.

This observation is in contrast to the behavior of entanglement for localization in 1D DTQW using disordered quantum coin operations. With spatially disordered coin operations, only a small decrease in entanglement is seen when compared to the entanglement due to standard DTQW\,\cite{Cha12} and with temporally and spatio-temporally disordered coin operations, enhancement of entanglement is seen\,\cite{Cha12, VAR13}. Though the states are localized, the degree of entanglement is not significantly affected because of the longer localization length of the disordered localized state when compared to the short localization length of topologically localized states. 
 
%======================
\section{Strength of the localized state in the presence of noise}
%======================

The application of noise to DTQWs is known to result in decoherence \cite{MK07, CSB07}, however small amounts of noise can also be advantageous for quantum algorithms and quantum transport.
Here we will look into the effect of $\sigma_x = \begin{bmatrix} \begin{array}{clcr}
  0      &     &   1
  \\  1 & &  0 \end{array} \end{bmatrix}$ (bit flip) noise on the topological quantum walk
 and show its effect on the localized and diffusive components. 
\par
The operation used for describing the two split-step DTQW evolution with $\sigma_x$ noise is given by
\bea
\rho(t) &=&  P \Big[ f_1   W(\theta_1, \theta_2) \rho(t-1) W(\theta_1,
\theta_2)^{\dagger} f_1^{\dagger} \Big ]  \nonumber \\
&&+  (1-P)  W(\theta_1, \theta_2)  \rho(t-1)  W(\theta_1, \theta_2)^{\dagger},
\label{sigmaX}
\eea
where $\rho(0) = |\Psi_\text{in}\rangle \langle \Psi_\text{in}|$, $W(\theta_1, \theta_2)$ is same as Eq.~\eqref{eq:4}, 
$f_1 \equiv \sigma_x \otimes  \mathbb{I}$ and $P$ is the magnitude of noise. No noise is described by $P=0$  and due to the fact that symmetries are not effected by bit flip noise, the maximum noise corresponds to $P=0.5$\,\cite{CSB07}.

  \begin{figure}[tb]
 \begin{center}
   \includegraphics[width=\linewidth]{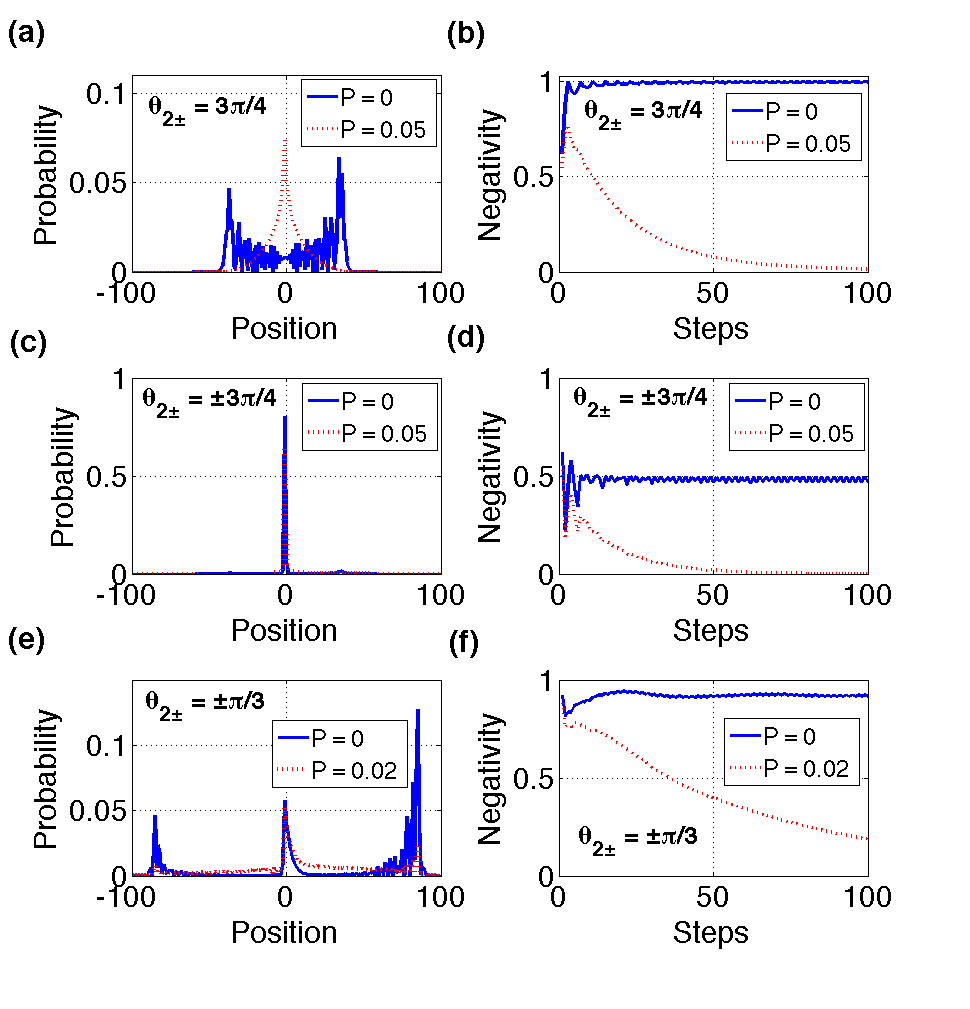}
   \caption{Effect of $\sigma_x$ noise on the split-step DTQW for different configurations of $\theta_{2\pm}$ when $\theta_1 = -\pi/4$.  (a), (c) and (e) show the probability distribution after 100 steps of walk in the absence and in the presence of noise. (b), (d) and (f) show the negativity as a function of the number of steps in absence and in presence of noise.  The probability distributions show that the noise affects the diffusing part of the walk significantly but hardly influences the topologically localized part.}
\label{sigmaxNoise}
 \end{center}
 \end{figure}
 
In Fig.~\ref{sigmaxNoise} we show the probability distributions and the corresponding values of negativity for the split-step DTQW for different configurations of $\theta_{2\pm}$ in the presence and absence of noise. Applying noise to an evolution which in the absence of noise leads to a delocalized state (see Fig.~\ref{sigmaxNoise}(a), blue line), now leads to a state that is located around the origin.  For a combination of $\theta_{2\pm}$ parameters resulting in a  probability distribution with both, localized and a diffusive components (see Fig.\,\ref{sigmaxNoise}(c) and (e)) the effect of noise results in a reduction of the probability for spreading in position space away from the origin, while the effect on the localized part is very small.  This indicates a robustness of topologically localized states to noise, which is absent for diffusive states.
This behaviour is also reflected in the negativity and in Figs.~\ref{sigmaxNoise}(b) and (d) and can see that the non-zero value of negativity in the absence of noise, indicative of a diffusive component in the probability distribution, decreases fast when noise is present. Even for noise levels as small as $P=0.02$,  the effect of noise on the delocalized probability distribution is very strong, see Fig.\,\ref{sigmaxNoise}(e).

\begin{figure}[tb]
  \centering
   \includegraphics[width=\linewidth]{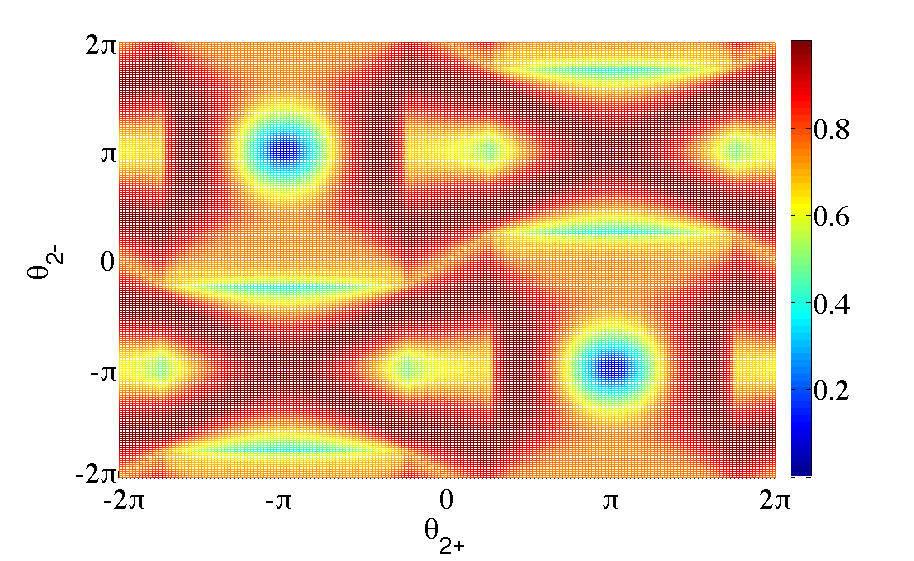}
    \includegraphics[width=\linewidth]{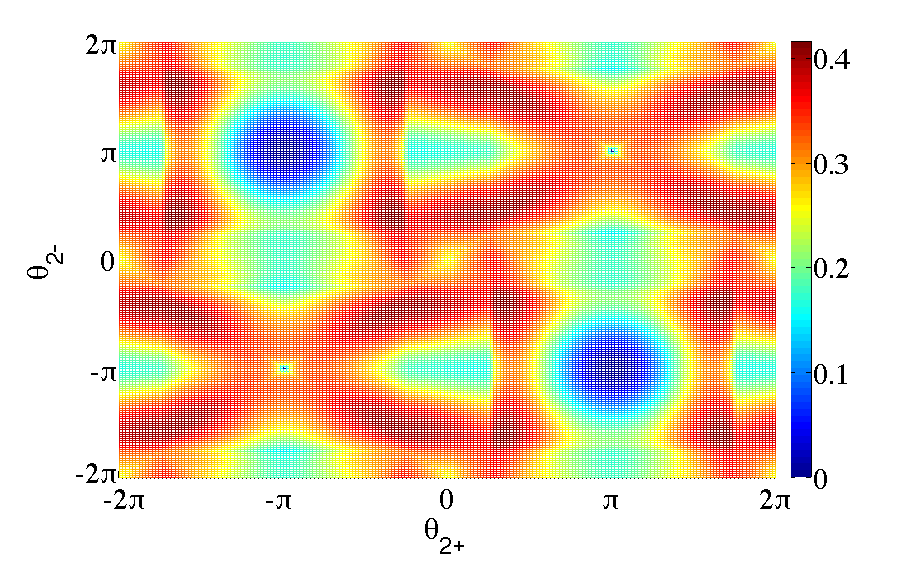}
   \caption{Negativity as a function of $\theta_{2-}$ and $\theta_{2+}$ for $\theta_{1} = -\pi/4$ for the the noiseless ($P=0.00$, upper) and the noisy ($P=0.02$, lower) split-step DTQW.  Note the difference in the color scale. 
   }
\label{noiseNegProfile} 
 \end{figure}

In Fig.\,\ref{noiseNegProfile} we show the negativity as function of $\theta_{2-}$ and $\theta_{2+}$ when $\theta_{1} = -\pi/4$ for evolutions without and with $\sigma_x$ noise. One can see that the effect of the noise results in a decrease of the overall negativity, but remains essentially unchanged in the regions where strongly localized states appear. This indicates the robustness of the topologically localized state against $\sigma_x$ noise.

\begin{figure}[tb]
   \includegraphics[width=\linewidth]{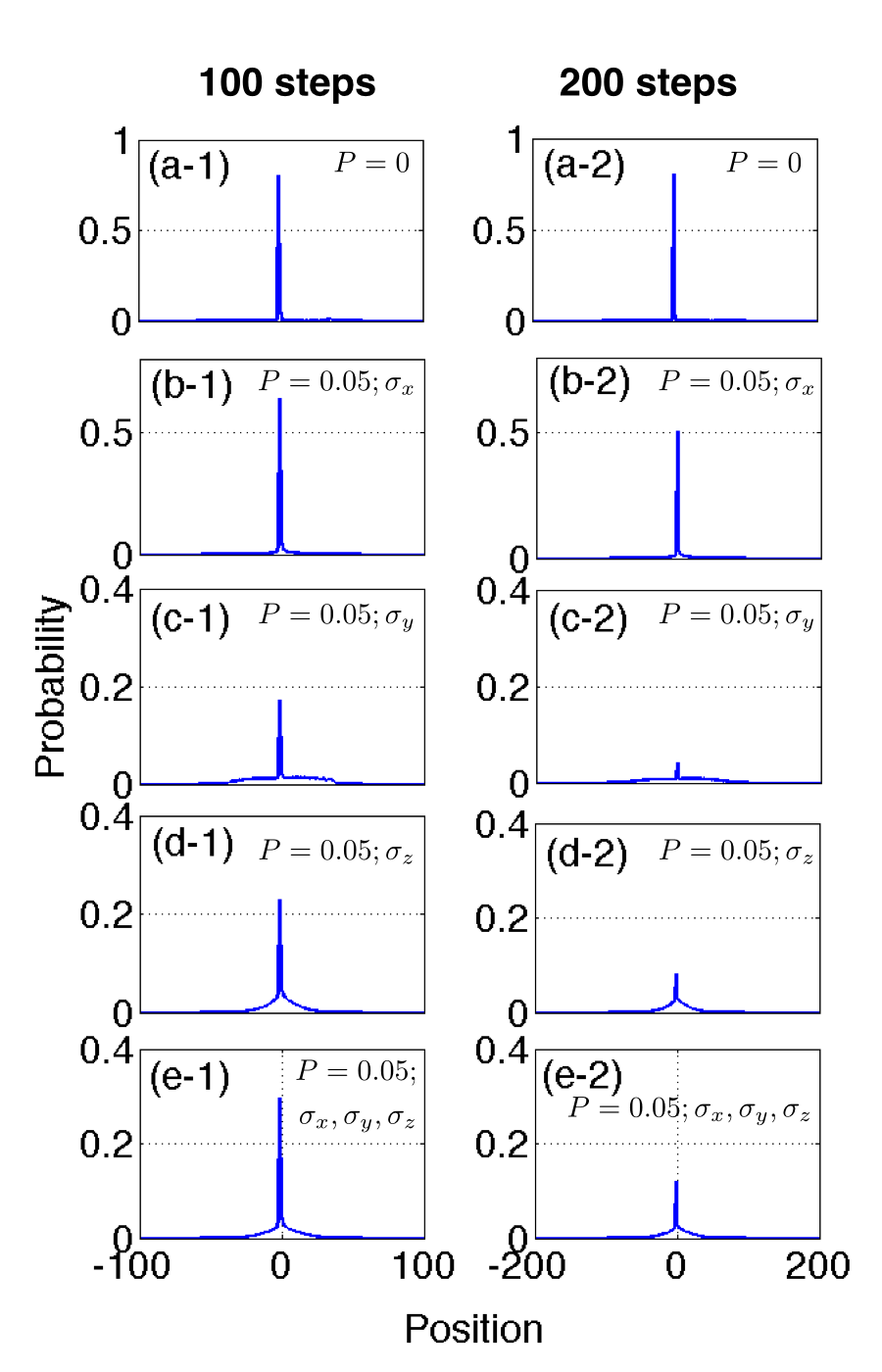}
   \caption{Effect of noise on topologically localized states of the split step DTQW with $\theta_1 = -\pi/4$, $\theta_{2-} = -3\pi/4$ and $\theta_{2+} = 3\pi/4$. The first and second column show the probability distribution after 100 step and 200 step of evolution, respectively. The first row is the probability distribution for evolution without noise ((a-1), (a-2)), the second row is for evolution with $\sigma_x$ noise ((b-1), (b-2)), the third row is for evolution with $\sigma_y$ noise ((c-1), (c-2)), the fourth row is for evolution with $\sigma_z$ noise ((d-1), (d-2)) and the fifth row is for evolution with depolarizing noise ((e-1), (e-2)). For all noisy evolutions the noise level is $P=0.05$ and one can clearly see that the localized state is robust against $\sigma_x$ but not other forms of noise, for which the localized component decays with increasing number of steps.}
\label{TQWnoise} 
 \end{figure}

However, topologically localized states in the split-step DTQW are not robust to other forms of noise. In Fig.~\ref{TQWnoise} we show the probability distributions for an evolution without (first row) and with noise (rest of the rows) after $100$ (first column) and $200$ (second column) steps. The $\sigma_{x}$ noise is the same as given in Eq.\,(\ref{sigmaX}) and the $\sigma_y = \begin{bmatrix} \begin{array}{clcr}
  0      &     &   -i
  \\  i & &  ~0 \end{array} \end{bmatrix}$ and $\sigma_z=\begin{bmatrix} \begin{array}{clcr}
  1      &     &   ~0
  \\  0 & &  -1 \end{array} \end{bmatrix}$ noises are obtained by replacing $\sigma_x$ by $\sigma_y$ and $\sigma_z$ in the Eq.~(\ref{sigmaX}). The operation to describe depolarizing noise is given by
\begin{align}
\rho(t) = \frac{1}{3}& \Big[ f_1   W(\theta_1, \theta_2) \rho(t-1) W(\theta_1,
\theta_2)^{\dagger} f_1^{\dagger}  \nonumber \\
& +f_2   W(\theta_1, \theta_2) \rho(t-1) W(\theta_1,
\theta_2)^{\dagger} f_2^{\dagger}  \nonumber \\ 
& +f_3   W(\theta_1, \theta_2) \rho(t-1) W(\theta_1,
\theta_2)^{\dagger} f_3^{\dagger} \Big ]  \nonumber \\
+&  (1-P)  W(\theta_1, \theta_2)  \rho(t-1)  W(\theta_1, \theta_2)^{\dagger}.
\label{sigmaDe}
\end{align}
Comparing the probability distributions of the localized states after evolution in the presence of these different kind of noises, one can clearly see that the topological states are robust only against $\sigma_x$ noise and a significant decrease in localized probability at the interface is visible for all other forms of noise. From Eqs.\,(\ref{eq:chiral symmetry operator}) and (\ref{eq:chiral edge states}) we can see that the edge states at $\varepsilon=0$ and $\pi$ are the eigenstates of the chiral symmetry operator $\Gamma$, which is identical to $f_1$ noise operator.
Therefore this symmetry is preserved in the presence of $f_1$ noise and the edge state protected.
However, this is not true for other forms of noise and the localized state can decay. 

\section{Conclusion}

Engineering  DTQWs with different combination of variable quantum coin and position shift operations allows to create a wide range of rich, topological phases. Choosing parameters $\theta_i$ with different topological numbers to the left and right side of an interface of the position space, leads to topology induced localized states, which are sometimes accompanied by a diffusing component. Identifying combinations resulting strong localization with minimal or completely absent diffusing components is important for simulations of artificial TIs and in this work we have shown that the negativity of a state can be used for such an identification. By exploring the negativity landscape as function of the quantum coin parameters we have linked the strength of the topologically localized states to the appearance of low values of the negativity. 

These topology induced localized states are different from the localized states originating from disordered DTQWs, where the presence of entanglement is usually robust against disorder. This therefore allows to differentiate between topologically localized states and localized state due to spatial and dynamic disorder in 1D DTQW. Finally, we have  demonstrated that the topologically localized component of a state is robust against $\sigma_x$ noise, whereas the diffusing component decays. We strongly believe that studies like this can lead to better engineering of the artificial materials to realize TIs.

\section*{Acknowledgments}
This work was supported by the Okinawa Institute of Science and Technology Graduate University.
H. O. was supported by Grant-in-Aid (Nos.\ 25800213 and 25390113) from the Japan Society for Promotion of Science.

%=================================

%\end{newlfm}

\begin{thebibliography}{99}
%=================================

\bibitem{Ria58+}  G. V. Riazanov, Sov. Phys., {\it The Feynman path integral for the Dirac equation}, Sov. Phys. JETP {\bf 6} 1107-1113 (1958)\,;  R. Feynman, \href{http://link.springer.com/article/10.1007\%2FBF01886518?LI=true} {Found. Phys. {\bf 16}, 507-531 (1986)}\,; K. R. Parthasarathy, \href{http://www.jstor.org/stable/3214153}{Journal of Applied Probability, {\bf 25}, 151-166 (1988)}. 

\bibitem{DM96} D.  A. Meyer,  {\it From quantum cellular automata to quantum lattice gases}, \href{http://link.springer.com/article/10.1007\%2FBF02199356}{J. Stat. Phys. {\bf 85},  551 (1996)}.

\bibitem{Amb03} A. Ambainis, \href{http://www.worldscientific.com/doi/abs/10.1142/S0219749903000383}{Int. J. Quantum. Inform. {\bf 01}, 507 (2003)}.

\bibitem{CCD+03} A. M. Childs, R. Cleve, E. Deotto, E. Farhi, S. Gutmann, and D. A. Spielman, {\it Exponential algorithmic speedup by quantum walk}, Proc. 35th ACM Symposium on Theory of Computing, pages 59-68 (2003).

\bibitem{Sze04} M. Szegedy. {\it Quantum speed-up of Markov chain based algorithms}, Foundations of Computer Science, 2004. Proceedings. 45th Annual IEEE Symposium on, pages 32-41 (2004).

\bibitem{AA05} S. Aaronson and A. Ambainis, {\it Quantum search of spatial regions}, Theory of Computing, 1(4):47-79 (2005).

\bibitem{MN07} F. Magniez and A. Nayak, {\it Quantum complexity of testing group commutativity}, Algorithmica, 48(3):221-232 (2007).

\bibitem{Amb07} A. Ambainis. {\it Quantum walk algorithm for element distinctness}, SIAM Journal on Computing, 37(1):210-239 (2007).

\bibitem{MNR+12} F. Magniez, A. Nayak, P. Richter, and M. Santha, {\it On the hitting times of quantum versus random walks}, Algorithmica, 63(1):91-116 (2012).

\bibitem{Chi09} A.M. Childs, {\it Universal computation by quantum walk}, \href{http://link.aps.org/doi/10.1103/PhysRevLett.102.180501}{Phys. Rev. Lett. {\bf 102}, 180501 (2009)}.

\bibitem{LCE10} N. B. Lovett, S. Cooper, M. Everitt, M. Trevers, and V. Kendon, {\it Universal quantum computation using the discrete-time quantum walk}, \href{http://link.aps.org/doi/10.1103/PhysRevA.81.042330}{Phys. Rev. A  {\bf 81}, 042330 (2010)}.

\bibitem{OKA05} T. Oka, N. Konno, R. Arita, and H. Aoki, {\it Breakdown of an Electric-Field Driven System: A Mapping to a Quantum Walk}, \href{http://link.aps.org/doi/10.1103/PhysRevLett.94.100602}{Phys. Rev. Lett. {\bf 94}, 100602 (2005)}. 

\bibitem{ECR07} Engel, G. S. {\it et al.}, {\it Evidence for wavelike energy transfer through quantum coherence in photosynthetic systems}, Nature {\bf 446}, 782-786 (2007).  

\bibitem{MRL08} Mohseni, M., Rebentrost, P., Lloyd, S. \& Aspuru-Guzik, A. {\it Environment-assisted quantum walks in photosynthetic energy transfer}, J. Chem. Phys. {\bf 129}, 174106 (2008). 

\bibitem{PH08}  Plenio, M. B. \& Huelga, S. F.  {\it Dephasing-assisted transport: quantum networks and biomolecules}, New J. Phys. {\bf 10}, 113019 (2008).

\bibitem{Str07} F. D. Strauch, {\it Relativistic effects and rigorous limits for discrete- and continuous-time quantum walks}, J. Math. Phys. {\bf 48}, 082102 (2007).

\bibitem{CBS10} C. M. Chandrashekar, S. Banerjee, \& R. Srikanth, {\it Relationship between quantum walks and relativistic quantum mechanics}, Phys. Rev. A  {\bf 81}, 062340 (2010).

\bibitem{GDB12} Giuseppe, D. M., Debbasch, F. \& Brachet, M. E., {\it Quantum walks as massless Dirac Fermion in curved space-time}, \href{http://dx.doi.org/10.1103/PhysRevA.88.042301}{Phys. Rev. A {\bf 88}, 042301 (2013)}. 

\bibitem{Cha13} C. M. Chandrashekar, {\it Two-component Dirac-like Hamiltonian for generating quantum walk on one-, two- and three-dimensional lattices}, \href{http://www.nature.com/srep/2013/131003/srep02829/full/srep02829.html} {Scientific Reports {\bf 3}, 2829 (2013)}. 
%\bibitem{Cha13} Chandrashekar, C. M.{\it  Two-component Dirac-like Hamiltonian for generating quantum walk on one-, two- and three-dimensional lattices. {\it Scientific Reports {\bf 3}, 2829 (2013)}. 

\bibitem{HK10} M. Z. Hasan and C. L. Kane, {\it Colloquium: Topological insulators}, \href{http://dx.doi.org/10.1103/RevModPhys.82.3045}{Rev. Mod. Phys. {\bf 82}, 3045 (2010)}.

\bibitem{QZ11} X.-L. Qi and S.-C. Zhang, {\it Topological insulators and superconductors}, \href{http://dx.doi.org/10.1103/RevModPhys.83.1057}{Rev. Mod. Phys. {\bf 83}, 1057 (2011)}.

\bibitem{Moo10} J. E. Moore, {\it The birth of topological insulators}, \href{http://www.nature.com/nature/journal/v464/n7286/full/nature08916.html}{Nature (London) {\bf 464}, 194-1998 (2010)}.


\bibitem{KRB10} T. Kitagawa, M. S. Rudner, E. Berg, and E. Demler,
{\it Exploring topological phases with quantum walks},
 Phys. Rev. A  {\bf 82},  033429 (2010).

\bibitem{OK11} H.\ Obuse and N.\ Kawakami, 
{\it Topological phases and delocalization of quantum walks in random
	environments}, 
Phys.\ Rev.\ B {\bf 84}, 195139 (2011).

\bibitem{KBF12} T. Kitagawa, M. A. Broome, A. Fedrizzi, M. S. Rudner, E. Berg, I. Kassal, A. Aspuru-Guzik, E. Demler, and A. G. White, {\it Observation of topologically protected bound states in photonic quantum walks}, \href{http://www.nature.com/ncomms/journal/v3/n6/full/ncomms1872.html} {Nature Communications {\bf 3}, 882 (2012)}.

\bibitem{Asb12} J. K. Asb\'oth, {\it Symmetries, topological phases, and bound states in the one-dimensional quantum walk}, \href{http://dx.doi.org/10.1103/PhysRevB.86.195414}{Phys. Rev. B {\bf 86}, 195414 (2012)}.


\bibitem{AO13} J. K. Asb\'oth  and H. Obuse,  
{\it Bulk-boundary correspondence for chiral symmetric quantum walks}
Phys. Rev. B  {\bf 88}, 121406(R) (2013).

\bibitem{TAD14} B.\ Tarasinski, J.\ K.\ Asb\'oth, and J.\ P.\ Dahlhaus,
{\it Scattering theory of topological phases in discrete-time quantum
	walks},
Phys.\ Rev.\ A {\bf 89}, 042327 (2014).


\bibitem{DLX03+} J. Du {\em et al.}, Phys. Rev. A {\bf 67}, 042316 (2003)\,; C. A.  Ryan {\em et al.}, \href{http://pra.aps.org/abstract/PRA/v72/i6/e062317}{Phys. Rev. A{\bf 72}, 062317 (2005)}\,; B. Do {\em et al.}, J. Opt. Soc. Am. B  {\bf 22}, 499 (2005)\,; H. B. Perets {\em et al.},  \href{http://prl.aps.org/abstract/PRL/v100/i17/e170506}{Phys. Rev. Lett. {\bf 100}, 170506 (2008)}\,; H. Schmitz {\em et al.}, \href{http://prl.aps.org/abstract/PRL/v103/i9/e090504}{Phys. Rev. Lett. {\bf 103}, 090504 (2009)}\,; F. Zahringer {\em et al.}, Phys. Rev. Lett. {\bf 104}, 100503 (2010)\,; K. Karski {\em et al.}, \href{http://www.sciencemag.org/content/325/5937/174.full}{Science  {\bf 325}, 174 (2009)}\,; A. Schreiber {\em et al.}, \href{http://prl.aps.org/abstract/PRL/v104/i5/e050502}{Phys. Rev. Lett., {\bf 104}, 050502 (2010)}\,; M. A. Broome {\em et al.}, \href{http://prl.aps.org/abstract/PRL/v104/i15/e153602}{Phys. Rev. Lett. {\bf 104}, 153602 (2010)}\,; A. Peruzzo {\em et al.}, 
 \href{http://www.sciencemag.org/content/329/5998/1500.full}{Science {\bf 329}, 1500 (2010)}\,; L. Sansoni {\em et al.}, Phys. Rev. Lett. {\bf 108}, 010502 (2012)\,; A. Schreiber  K. N. Cassemiro, V. Potocek, A. G\'abris, I. Jex, and Ch. Silberhorn, %{\it Decoherence and Disorder in Quantum Walks: From Ballistic Spread to Localization}, 
 \href{http://link.aps.org/doi/10.1103/PhysRevLett.106.180403}{Phys. Rev. Lett. {\bf 106}, 180403 (2011)}\,;A. Crespi {\em et al.},% {\it Anderson localization of entangled photons in an integrated quantum walk},
 \href{http://www.nature.com/nphoton/journal/v7/n4/full/nphoton.2013.26.html}{Nat. Phot. {\bf 7}, 323 (2013)}\,; J. D. A. Meinecke {\em et al.}, %{\it Coherent Time Evolution and Boundary Conditions of Two-Photon Quantum Walks}, 
 \href{http://dx.doi.org/10.1103/PhysRevA.88.012308}{Phys. Rev. A {\bf 88}, 012308 (2013)}.


\bibitem{Schnyder08} 
A.\ P.\ Schnyder, S.\ Ryu, A.\ Furusaki, and A.\ W.\ W.\ Ludwig,
{\it Classification of topological insulators and superconductors in three spatial dimensions
}
Phys.\ Rev.\ B, 
{\bf 78}, 195125 (2008).


\bibitem{Cha12} C. M. Chandrashekar, arXiv:1212.5984 (2012).

\bibitem{VAR13} R. Vieira, E. P. M. Amorim, and G. Rigolin, Phys. Rev. Lett. {\bf 111}, 180503 (2013).


\bibitem{MK07} O. Maloyer and V. Kendon, {\it Decoherence versus entanglement in coined quantum walks}, \href{http://iopscience.iop.org/1367-2630/9/4/087/}{New J. Phys. {\bf 9}, 87  (2007)}. 

\bibitem{GC10} S. K. Goyal and C. M. Chandrashekar, {\it Spatial entanglement using a quantum walk on a many-body system}, \href{http://iopscience.iop.org/1751-8121/43/23/235303}{J. Phys. A: Math. Theor. {\bf 43}, 235303 (2010)}.

\bibitem{CSB07}C. M. Chandrashekar, R. Srikanth, and S. Banerjee, {\it Symmetries and noise in quantum walk}, Phys. Rev. A, {\bf 76}, 022316 (2007). 


\end{thebibliography}
\end{document}